\documentclass{article}

\usepackage[utf8]{inputenc}
\usepackage{amsmath, amsfonts, dsfont}
\usepackage{graphicx}

\usepackage[numbers,square]{natbib}

\usepackage{geometry}
\title{Computing trading strategies based on financial sentiment data using
evolutionary optimization}

\author{
Ronald Hochreiter
}

\date{April 2015}

\begin{document}

\maketitle

\begin{abstract}
\noindent In this paper we apply evolutionary optimization techniques to compute
optimal rule-based trading strategies based on financial sentiment data.
The sentiment data was extracted from the social media service
StockTwits to accommodate the level of bullishness or bearishness of the
online trading community towards certain stocks. Numerical results for
all stocks from the Dow Jones Industrial Average (DJIA) index are
presented and a comparison to classical risk-return portfolio selection
is provided.
\end{abstract}

\section{Introduction}\label{introduction}

In this paper we apply evolutionary optimization techniques to compute
optimal rule-based trading strategies based on financial sentiment data.
The number of application areas in the field of sentiment analysis is
huge, see especially \cite{feldman2013techniques} for a comprehensive
overview. The field of Finance attracted research on how to use specific
financial sentiment data to find or optimize investment opportunities
and strategies, see e.g. \cite{bollen2011twitter},
\cite{oliveira2013predictability}, and \cite{smailovic2014stream}.

This paper is organized as follows. Section \ref{financial-sentiments}
describes the financial sentiment data used for the evolutionary
approach to optimize trading strategies and portfolios. Section
\ref{evolutionary-investment-strategy-generation} presents an
evolutionary optimization algorithm to create optimal trading strategies
using financial sentiment data and how to build a portfolio using
single-asset trading strategies. Section \ref{numerical-results}
contains numerical results obtained with the presented algorithm and a
comparison to classical risk-return portfolio optimization strategies as
proposed by \cite{markowitz1952portfolio} using stock market data from
all stocks in the Dow Jones Industrial Average (DJIA) index. Section
\ref{conclusion} concludes the paper.

\section{Financial Sentiments}\label{financial-sentiments}

We apply financial sentiment data created by
PsychSignal\footnote{http://www.psychsignal.com/}. The PsychSignal
technology utilizes the wisdom of crowds in order to extract meaningful
analysis, which is not achievable through the study of single
individuals, see \cite{gottschalk1969} for a general introduction to
measurement of psychological states through verbal behavior. Let a group
of individuals together be a crowd. Not all crowds are wise, however
four elements have been identified, which are required to form a wise
crowd: diversity of opinion, independence, decentralization and
aggregation as proposed by \cite{Surowiecki2005}. These four elements
are sometimes present in some forms of social media platforms, e.g.~in
the financial community StockTwits, from which the crowd wisdom used for
the evolutionary approach described in this paper is derived.

Emotions are regarded as being unique to individual persons and
occurring over brief moments in time. Let a mood be a set of emotions
together. In order to quantify the collective mood of a crowd, distinct
emotions of individual members within the crowd must be quantified.
Subsequently, individual emotions can be aggregated to form a collective
crowd mood. PsychSignals' Natural Language Processing Engine is tuned to
the social media language of individual traders and investors based on
the general findings of e.g. \cite{das2007yahoo} and of
\cite{tumarkin2001news} for the financial domain. The engine further
targets and extracts emotions and attitudes in that distinct language
and categorizes as well as quantifies these emotions from text. The
methodology is based on the linguistic inquiry and word count (LIWC)
project, which is available publicly\footnote{http://www.liwc.net/}. See
also \cite{oliveira2014automatic} for a description of an algorithm on
how to generate such a semantic lexicon for financial sentiment data
directly.

The main idea is to assign a degree of bullishness or bearishness on
stocks depending on the messages, which are sent through
StockTwits\footnote{http://www.stocktwits.com/}, which utilizes
Twitter's application programming interface (API) to integrate
StockTwits as a social media platform of market news, sentiment and
stock-picking tools. StockTwits utilized so called \emph{cashtags} with
the stock ticker symbol, similar to the Twitter \emph{hashtag}, as a way
of indexing people's thoughts and ideas about companies and their
respective stocks. The available sentiment data format is described in
Tab. \ref{tab:psychsignal}. The data was obtained through
Quandl\footnote{http://www.quandl.com/}, where PsychSignal's sentiment
data for stocks can be accessed easily.

\begin{table}
\caption{PsychSignal.com StockTwits sentiment data format per asset.}
\label{tab:psychsignal}

\centering 
\begin{tabular}{ll}
\hline
Variable & Content \\
\hline
Date & Day of the analyzed data. \\
\(I_{\text{bull}}\) & Each message's language 
strength of bullishness on a 0-4 scale. \\
\(I_{\text{bear}}\) & Each message's language
strength of bearishness on a 0-4 scale. \\
\(n_{\text{bull}}\) & Total count of bullish sentiment
messages. \\
\(n_{\text{bear}}\) & Total count of bearish sentiment
messages. \\
\(n_{\text{total}}\) & Total number of messages. \\
\hline
\end{tabular}

\end{table}

Both intensities $I_{\text{bull}}$ and $I_{\text{bear}}$ are measured on
a real-valued scale from $0$ to $4$, where $0$ means no bullish/bearish
sentiment and $4$ the strongest bullish/bearish sentiment. We normalize
these values to $1$ by diving the respective value by $4$ and obtain the
variables $i_{\text{bull}}$ and $i_{\text{bear}}$. Furthermore, we
create two relative variables for the number of bullish and bearish
messages, i.e. $r_{\text{bull}} = n_{\text{bull}} / n_{\text{total}}$ as
well as $r_{\text{bear}} = n_{\text{bear}} / n_{\text{total}}$, such
that we end up in the final data format we are going to use for
subsequent analysis. See Tab. \ref{tab:converted} for an example of the
stock with the ticker symbol BA (The Boeing Company).

\begin{table}
\caption{Sentiment values for stock BA starting at the first trading days in 2011.}
\label{tab:converted}

\centering
\begin{tabular}{lrrrrrr}
  \hline
 & $i_{\text{bull}}$ & $i_{\text{bear}}$ & $r_{\text{bull}}$ & $r_{\text{bear}}$ & $n_{\text{total}}$ \\ 
  \hline
2011-01-03 & 0.59 & 0 & 0.50 & 0 & 4 \\ 
2011-01-04 & 0 & 0 & 0 & 0 & 1 \\ 
2011-01-05 & 0 & 0.11 & 0 & 1 & 1 \\ 
2011-01-06 & 0.61 & 0 & 0.25 & 0 & 4 \\ 
2011-01-07 & 0.52 & 0 & 0.17 & 0 & 6 \\ 
2011-01-11 & 0.67 & 0 & 1 & 0 & 2 \\ 
   \hline
\end{tabular}
\end{table}

\section{Evolutionary Investment Strategy
Generation}\label{evolutionary-investment-strategy-generation}

We aim at creating an evolutionary optimization approach to generate
optimal trading strategies for single stocks based on the sentiment
analysis data described above. Evolutionary and Genetic Programming
techniques have been applied to various financial problems successfully.
See especially the series of books on Natural Computing in Finance for
more examples, i.e. \cite{ncfin2008}, \cite{ncfin2009}, and
\cite{ncfin2010}. Generating automatic trading rules has been a core
topic in this domain, see especially \cite{bradley2010evolving},
\cite{brabazon2004evolving}, \cite{brabazon2006intra},
\cite{Lipinski2004}, and the references therein.

One main technique in the field of meta-heuristics and technical trading
is to let the optimizer generate optimal investment rules given a set of
technical indicators. However, instead of using a variety of technical
indicators for generating an optimal trading rule, we use the above
described financial sentiment data to create investment rules. Thereby
we start by using a simplified rule-set approach, whereby the rules are
generated by a special genotype encoding. Furthermore, as we are
considering to create a portfolio allocation out of the single asset
strategies and additionally focus on stocks only, we do not allow for
shorting assets, i.e.~the decision is whether to enter or exit a long
position on a daily basis. The rule is based on the respective sentiment
values, such that this basic rule-set can be defined as shown in Eq.
(\ref{equ:rulesetgeneral}).

\begin{eqnarray}
\begin{array}{l}
\big [ \text{IF}(i_{\text{bull}} \geq v_1) \big ]_{b_1} \big [ \text{AND} \big ]_{b_1 \& b_2} \big [ \text{IF} (r_{\text{bull}} \geq v_2) \big ]_{b_2} \text{THEN long position}. \\
\big [ \text{IF}(i_{\text{bear}} \geq v_3) \big ]_{b_3} \big [ \text{AND} \big ]_{b_3 \& b_4} \big [ \text{IF} (r_{\text{bear}} \geq v_4) \big ]_{b_4} \text{THEN exit position}. \\
\end{array}
\label{equ:rulesetgeneral}
\end{eqnarray}

Each chromosome within the evolutionary optimization process consists of
the values $$(b_1, b_2, b_3, b_4, v_1, v_2, v_3, v_4),$$ where the $b$
values are binary encoded (0, 1) and the $v$ values are real values
between $0$ and $1$. The $b$ values indicate whether the respective part
of the rule notated in square brackets is included (1) or not (0), while
the $v$ values represent the concrete values within the conditions.
Consider the following example: the (randomly chosen) chromosome
(0,1,1,1,0.4,0.3,0.5,0.2) results in the rule-set shown in Eq.
(\ref{equ:rulesetexample}).

\begin{eqnarray}
\begin{array}{l}
\text{IF } (r_{\text{bull}} \geq 0.3) \text{ THEN long position}. \\
\text{IF } (i_{\text{bear}} \geq 0.5) \text{ AND } \text{IF } (r_{\text{bear}} \geq 0.2) \text{ THEN exit position}. \\
\end{array}
\label{equ:rulesetexample}
\end{eqnarray}

In this special case, the sum of $b_1$ and $b_2$ as well as $b_3$ and
$b_4$ must be greater or equal to $1$, to have at least one condition
for entering and leaving the long position. We end up with nine
different possible assignments for $b$. A repair operator has to be
applied after each evolutionary operation, which may distort this
structure.

The evaluation of the chromosomes is such that the respective trading
strategy is tested on the in-sample testing set of length $T$, i.e.~we
obtain a series of returns $r_1, \ldots, r_T$ for each chromosome, which
can be evaluated with different financial metrics. The following
strategy performance characteristics are considered:

\begin{itemize}
\itemsep1pt\parskip0pt\parsep0pt
\item
  The cumulative return $r$, and the standard deviation $\sigma$.
\item
  The maximum drawdown $d$, and the Value-at-Risk $v_\alpha$
  $(\alpha=0.05)$, as well as
\item
  the ratio $s$ of expected return divided by the standard deviation,
  which is based on the Sharpe-ratio proposed by
  \cite{sharpe1994sharpe}.
\end{itemize}

We use simple mutation operators for new populations because the
chromosome encoding of the investment rule described above is short,
i.e.~contains only eight genes. The following mutation operators are
applied:

\begin{itemize}
\itemsep1pt\parskip0pt\parsep0pt
\item
  $b$ binary flip: One randomly selected gene of the binary $b$ part is
  $0-1$ flipped. The resulting chromosome needs to be repaired with the
  repair operator, which itself determines randomly, which of the two
  possibilities is set to $1$ if necessary.
\item
  $v$ random mutation: One randomly selected gene of the binary $v$ part
  is replaced by a uniform random variable between $0$ and $1$.
\item
  $v$ mutation divided in half: One randomly selected gene of the binary
  $v$ part is divided in half. The rationale of this operation is that
  the intensities of bullishness and bearishness are often small, see
  e.g.~Tab. \ref{tab:statistics} for the statistics of the sentiment
  values for a selected stock.
\end{itemize}

\begin{table}
\caption{Statistical summary of sentiment values for stock BA 2010-2014.}
\label{tab:statistics}

\centering
\begin{tabular}{lllllll}

\hline
& Minimum & First Quantile & Median & Mean & Third Quantile & Maximum \\ \hline
$i_{bull}$ & 0 & 0 & 0.3821 & 0.2987 & 0.5050 & 0.8250 \\
$i_{bear}$ & 0 & 0 & 0 & 0.1763 & 0.3887 & 0.86 \\ \hline

\end{tabular}
\end{table}

Besides these operators, elitist selection is applied as well as a
number of random additions will be added to each new population. The
structure of the algorithm is a general genetic algorithm, see e.g.
\cite{blum2003metaheuristics} for a description of this class of
meta-heuristics.

The analysis above is based on single assets. To compose a portfolio out
of the single investment strategies, the resulting portfolio will be
created as an equally weighted representation of all assets, which are
currently selected to be in a long position by its respective trading
strategy for each day.

\section{Numerical Results}\label{numerical-results}

In this section we begin with a description of the data used to compute
numerical results in Section \ref{data}. Section
\ref{results-of-the-evolutionary-optimization} summarizes the in-sample
and out-of-sample results of the evolutionary sentiment trading
strategy. A short overview of classical risk-return portfolio
optimization is given in Section \ref{classical-portfolio-optimization},
and finally a performance comparison is presented in Section
\ref{performance-comparison}. Everything was implemented using the
statistical computing language R \cite{R2014}.

\subsection{Data}\label{data}

We use data from all stocks from the Dow Jones Industrial Average (DJIA)
index using the composition of September 20, 2013, i.e.~using the stocks
with the ticker symbols AXP, BA, CAT, CSCO, CVX, DD, DIS, GE, GS, HD,
IBM, INTC, JNJ, JPM, KO, MCD, MMM, MRK, MSFT, NKE, PFE, PG, T, TRV, UNH,
UTX, V, VZ, WMT, XOM.

Training data is taken from the beginning of 2010 until the end of 2013.
The out-of-sample tests are applied to data from the year 2014.

\subsection{Results of the Evolutionary
Optimization}\label{results-of-the-evolutionary-optimization}

For each stock, the optimal strategy was computed. The evolutionary
parameters were set to be as follows:

\begin{itemize}
\itemsep1pt\parskip0pt\parsep0pt
\item
  The initial population size has been set to $100$, and each new
  population
\item
  contains the $10$ best chromosomes of the previous population (elitist
  selection), as well as
\item
  $20$ of each of the three mutation operators described above, and
\item
  $10$ random chromosomes, such that the population size is $80$.
\end{itemize}

For evaluation purposes, the parameter $s$ will be maximized. Of course,
the system is flexible to use any other risk metric or a combination of
metrics. See Tab. \ref{tab:evoresults} for the in-sample performance
results comparing a long-only buy-and-hold strategy of each asset
compared to the trading strategy of the best respective strategy,
e.g.~the best strategy for AXP is $(1,1,1,0,0.44,0.41,0.41,0.17)$ and
for BA $(1,0,1,0,0.41,0.37,0.5,0.41)$, while for CAT it is
$(0,1,1,1,0.195,0.34,0.02,0.24)$ to give an impression of single
strategy results. The cumulative return performance $r$ is raised
(sometimes significantly) for almost all assets except for MCD, UTX, V.
However, in those three cases the decrease in profit is low. The
standard deviation $\sigma$ is lower (i.e.~better) in all cases, which
was expected as the algorithm leaves the long-position for a certain
time, such that the standard deviation clearly has to decrease. The
Sharpe-ratio like metric $s$ is better for all assets but DIS, JNJ, UTX,
XOM. Again, the loss in all four cases is low compared to the gain of
the other positions. In summary, the in-sample results show that the
fitting of the algorithm works very well.

\begin{table}
\caption{Single stock in-sample results of the evolutionary optimization.}
\label{tab:evoresults}

\centering
\begin{tabular}{rrrrrrrr}
  \hline
& \multicolumn{3}{l}{Long-only stock} & \hspace{1em} & \multicolumn{3}{l}{Trading strategy} \\
 & $r$ & $\sigma$ & $s$ & & $r$ & $\sigma$ & $s$ \\ 
  \hline
  AXP & 1.223 & 0.016 & 0.057   & & 1.574 & 0.013 & 0.068 \\ 
  BA & 1.450 & 0.016 & 0.063    & & 1.771 & 0.013 & 0.070 \\ 
  CAT & 0.575 & 0.018 & 0.034   & & 0.978 & 0.014 & 0.043 \\ 
  CSCO & -0.070 & 0.019 & 0.006 & & 1.246 & 0.011 & 0.061 \\ 
  CVX & 0.597 & 0.013 & 0.042   & & 0.723 & 0.012 & 0.044 \\ 
  DD & 0.912 & 0.015 & 0.051    & & 1.177 & 0.011 & 0.070 \\ 
  DIS & 1.351 & 0.015 & 0.065   & & 1.522 & 0.013 & 0.065 \\ 
  GE & 0.842 & 0.015 & 0.048    & & 1.054 & 0.013 & 0.050 \\ 
  GS & 0.042 & 0.019 & 0.012    & & 0.662 & 0.010 & 0.041 \\ 
  HD & 1.825 & 0.014 & 0.082    & & 2.209 & 0.012 & 0.087 \\ 
  IBM & 0.430 & 0.012 & 0.036   & & 0.711 & 0.005 & 0.083 \\ 
  INTC & 0.249 & 0.015 & 0.022  & & 0.600 & 0.009 & 0.043 \\ 
  JNJ & 0.415 & 0.008 & 0.045   & & 0.415 & 0.008 & 0.041 \\ 
  JPM & 0.399 & 0.019 & 0.027   & & 0.873 & 0.016 & 0.037 \\ 
  KO & -0.277 & 0.019 & -0.004  & & 0.363 & 0.006 & 0.042 \\ 
  MCD & 0.549 & 0.009 & 0.052   & & 0.515 & 0.006 & 0.060 \\ 
  MMM & 0.688 & 0.013 & 0.048   & & 0.795 & 0.012 & 0.051 \\ 
  MRK & 0.359 & 0.012 & 0.032   & & 0.516 & 0.010 & 0.041 \\ 
  MSFT & 0.222 & 0.014 & 0.021  & & 0.509 & 0.012 & 0.031 \\ 
  NKE & 0.190 & 0.022 & 0.022   & & 0.511 & 0.019 & 0.030 \\ 
  PFE & 0.677 & 0.012 & 0.048   & & 0.805 & 0.011 & 0.049 \\ 
  PG & 0.332 & 0.009 & 0.036    & & 0.406 & 0.007 & 0.046 \\ 
  T & 0.238 & 0.010 & 0.026     & & 0.397 & 0.007 & 0.040 \\ 
  TRV & 0.805 & 0.012 & 0.053   & & 0.946 & 0.011 & 0.064 \\ 
  UNH & 1.400 & 0.016 & 0.063   & & 2.443 & 0.013 & 0.091 \\ 
  UTX & 0.621 & 0.013 & 0.042   & & 0.563 & 0.011 & 0.042 \\ 
  V & 1.530 & 0.017 & 0.062     & & 1.494 & 0.010 & 0.072 \\ 
  VZ & 0.471 & 0.011 & 0.041    & & 0.572 & 0.009 & 0.045 \\ 
  WMT & 0.464 & 0.009 & 0.045   & & 0.654 & 0.007 & 0.058 \\ 
  XOM & 0.473 & 0.012 & 0.039   & & 0.506 & 0.010 & 0.036 \\ 
   \hline
\end{tabular}
\end{table}

\subsection{Classical Portfolio
Optimization}\label{classical-portfolio-optimization}

To compare the performance of the portfolio created with single asset
investment strategies based on financial sentiments with a standard
approach to portfolio optimization, we construct a portfolio using
classical risk-return portfolio selection techniques.
\cite{markowitz1952portfolio} pioneered the idea of risk-return optimal
portfolios using the standard deviation of the portfolios profit and
loss function as risk measure. In this case, the optimal portfolio $x$
is computed by solving the quadratic optimization problem shown in Eq.
\ref{f:marko1}. The investor needs to estimate a vector of expected
returns $r$ of the assets under consideration as well as the covariance
matrix $\mathbb{C}$. Finally the minimum return target $\mu$ has to be
defined. Any standard quadratic programming solver can be used to solve
this problem numerically.

\begin{eqnarray}
\begin{array}{ll}
\text{minimize} & x^T \mathbb{C} x \\
\text{subject to} & r \times x \geq \mu \\
& \sum x = 1 \\
\label{f:marko1}
\end{array}
\end{eqnarray}

In addition, we also compare the performance to the 1-over-N portfolio,
which equally weights every asset under consideration. It has been shown
that there are cases, where this simple strategy outperforms clever
optimization strategies, see e.g. \cite{demiguel2009optimal}.

\subsection{Performance Comparison}\label{performance-comparison}

\begin{figure}

\centering
\includegraphics[width=12cm]{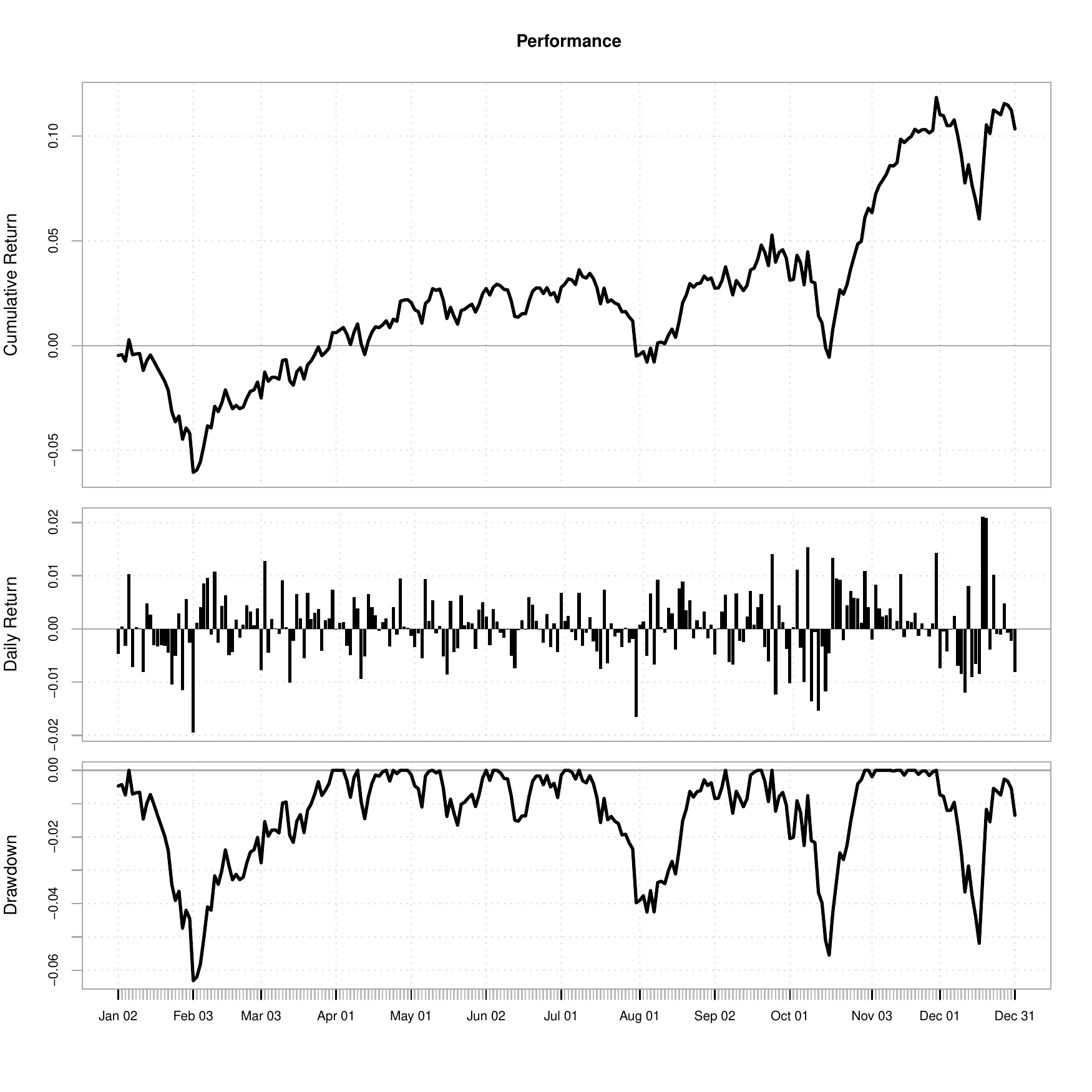}

\caption{Out-of-sample performance of a buy-and-hold Markowitz portfolio in 2014.}
\label{f:performace-markowitz}
\end{figure}

\begin{figure}

\centering
\includegraphics[width=12cm]{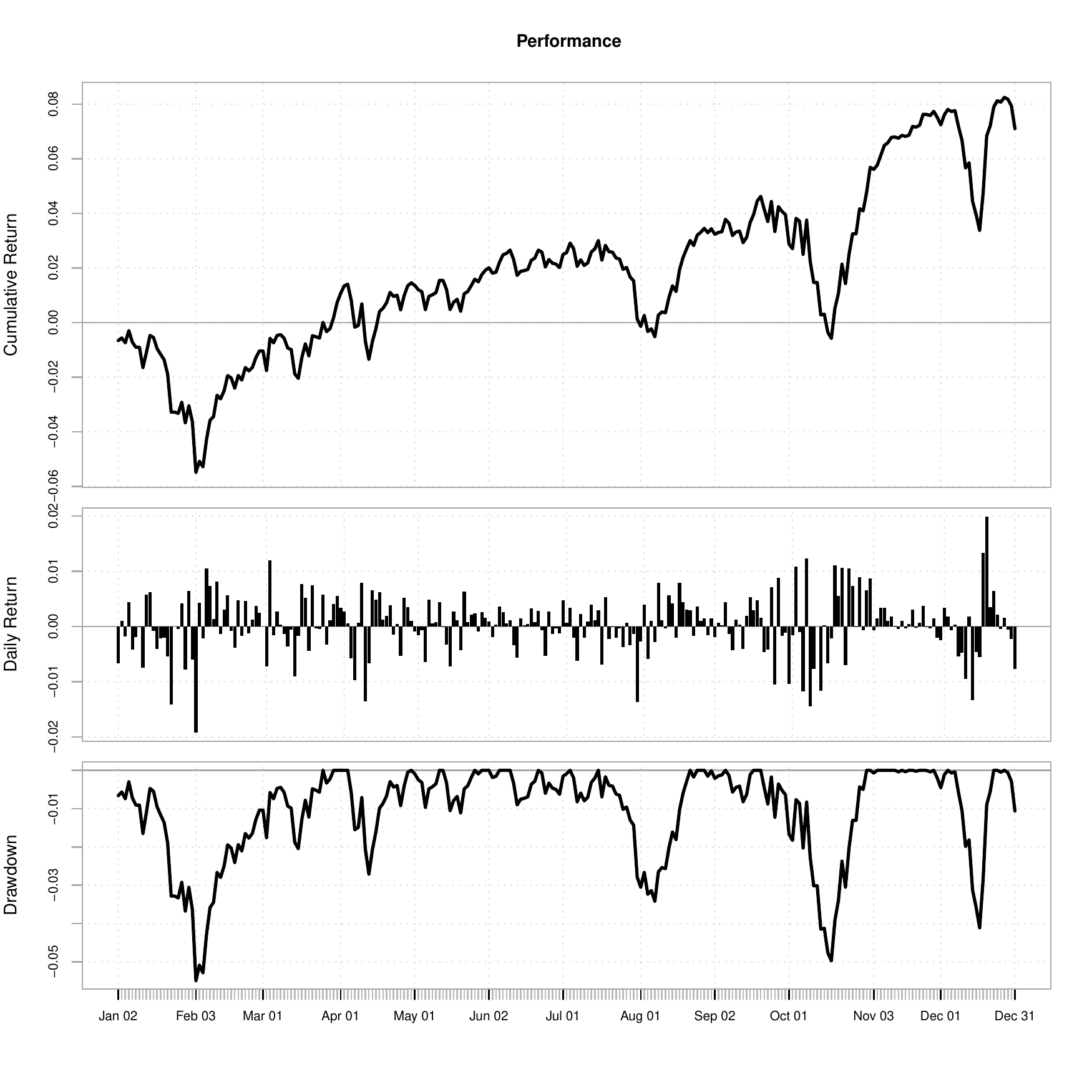}

\caption{Out-of-sample performance of an equally weighted portfolio out of the evolutionary sentiment trading strategies in 2014.}
\label{f:performace-evo}
\end{figure}

The asset composition of the optimal Markowitz portfolio is shown in
Tab. \ref{tab:markowitz} - only eight out of the $30$ assets are
selected. The underlying covariance matrix was estimated from daily
returns of the training data, i.e.~using historical returns from the
beginning of 2010 until the end of 2013. This portfolio is used as a
buy-and-hold portfolio over the year 2014. This out-of-sample
performance is shown in Fig.
\ref{f:performace-markowitz}\footnote{Performance graphs are generated using the {\tt PerformanceAnalytics} R package \cite{peterson2014}.}.
While the performance of the 1-over-N portfolio is not shown
graphically, Fig. \ref{f:performace-evo} depicts the performance of a
portfolio, which is created by equally weighting all single asset
trading strategies computed by the evolutionary optimization algorithm
based on financial sentiment data into one portfolio. To get a better
impression of the differences see Tab. \ref{tab:riskresult}, where some
important risk metrics are summarized for all three strategies. The
evolutionary trading portfolio exhibits better risk properties than both
other portfolios in all five metrics. Especially important is the
reduction of the maximum drawdown, which is of importance to asset
managers nowadays, because investors are increasingly looking to this
metric if they are searching for secure portfolios.

\begin{table}

\caption{Optimal Markowitz portfolio using daily return data from 2010-2013.}
\label{tab:markowitz}

\centering
\begin{tabular}{lrrrrrrrr}

\hline

Ticker symbol &    HD &   JNJ &   MCD &    PG &   UNH &     V &    VZ &   WMT \\ 
Portfolio weight [\%] & 10.26 & 16.69 & 22.67 & 11.92 &  6.41 &  4.22 &  7.56 & 20.27 \\ 

\hline
\end{tabular}

\end{table}

\begin{table}

\caption{Selected risk metrics for the different out-of-sample tests.}
\label{tab:riskresult}

\centering
\begin{tabular}{rrrr}
  \hline
& Markowitz & 1-over-N & Evolutionary \\ 
  \hline
  Semi Deviation & 0.0042 & 0.0048 & 0.0038 \\ 
  Downside Deviation (Rf=0\%) & 0.0040 & 0.0047 & 0.0037 \\ 
  Maximum Drawdown & 0.0631 & 0.0687 & 0.0549 \\ 
  Historical VaR (95\%) & -0.0092 & -0.0105 & -0.0081 \\ 
  Historical ES (95\%) & -0.0125 & -0.0157 & -0.0124 \\ 
   \hline
\end{tabular}
\end{table}

\section{Conclusion}\label{conclusion}

In this paper an evolutionary optimization approach to compute optimal
rule-based trading strategies based on financial sentiment data has been
developed. It can be shown that a portfolio composed out of the single
trading strategies outperforms classical risk-return portfolio
optimization approaches in this setting. The next step is to include
transaction costs to see how this active evolutionary strategy loses
performance when transaction costs are considered. Future extensions
include extensive numerical studies on other indices as well as using
and comparing different evaluation risk metrics or a combination of
metrics. One may also consider to create a more flexible rule-generating
algorithm e.g.~by using genetic programming. Finally, to achieve an even
better out-of-sample performance the recalibrating of the trading
strategy can be done using a rolling horizon approach every month.


\begin{thebibliography}{22}
\providecommand{\natexlab}[1]{#1}
\providecommand{\url}[1]{\texttt{#1}}
\expandafter\ifx\csname urlstyle\endcsname\relax
  \providecommand{\doi}[1]{doi: #1}\else
  \providecommand{\doi}{doi: \begingroup \urlstyle{rm}\Url}\fi

\bibitem[Blum and Roli(2003)]{blum2003metaheuristics}
C.~Blum and A.~Roli.
\newblock Metaheuristics in combinatorial optimization: Overview and conceptual
  comparison.
\newblock \emph{ACM Computing Surveys}, 35\penalty0 (3):\penalty0 268--308,
  2003.

\bibitem[Bollen et~al.(2011)Bollen, Mao, and Zeng]{bollen2011twitter}
J.~Bollen, H.~Mao, and X.~Zeng.
\newblock Twitter mood predicts the stock market.
\newblock \emph{Journal of Computational Science}, 2\penalty0 (1):\penalty0
  1--8, 2011.

\bibitem[Brabazon and O'Neill(2008)]{ncfin2008}
A.~Brabazon and M.~O'Neill, editors.
\newblock \emph{Natural Computing in Computational Finance}, volume 100 of
  \emph{Studies in Computational Intelligence}.
\newblock Springer, 2008.

\bibitem[Brabazon and O'Neill(2009)]{ncfin2009}
A.~Brabazon and M.~O'Neill, editors.
\newblock \emph{Natural Computing in Computational Finance, Volume 2}, volume
  185 of \emph{Studies in Computational Intelligence}.
\newblock Springer, 2009.

\bibitem[Brabazon and O’Neill(2004)]{brabazon2004evolving}
A.~Brabazon and M.~O’Neill.
\newblock Evolving technical trading rules for spot foreign-exchange markets
  using grammatical evolution.
\newblock \emph{Computational Management Science}, 1\penalty0 (3-4):\penalty0
  311--327, 2004.

\bibitem[Brabazon and O’Neill(2006)]{brabazon2006intra}
A.~Brabazon and M.~O’Neill.
\newblock Intra-day trading using grammatical evolution.
\newblock In A.~Brabazon and M.~O’Neill, editors, \emph{Biologically Inspired
  Algorithms for Financial Modelling}, pages 203--210. Springer, 2006.

\bibitem[Brabazon et~al.(2010)Brabazon, O'Neill, and Maringer]{ncfin2010}
A.~Brabazon, M.~O'Neill, and D.~Maringer, editors.
\newblock \emph{Natural Computing in Computational Finance, Volume 3}, volume
  293 of \emph{Studies in Computational Intelligence}.
\newblock Springer, 2010.

\bibitem[Bradley et~al.(2010)Bradley, Brabazon, and
  O’Neill]{bradley2010evolving}
R.~G. Bradley, A.~Brabazon, and M.~O’Neill.
\newblock Evolving trading rule-based policies.
\newblock \emph{Lecture Notes in Computer Science}, 6025:\penalty0 251--260,
  2010.

\bibitem[Das and Chen(2007)]{das2007yahoo}
S.~R. Das and M.~Y. Chen.
\newblock Yahoo! for {A}mazon: Sentiment extraction from small talk on the web.
\newblock \emph{Management Science}, 53\penalty0 (9):\penalty0 1375--1388,
  2007.

\bibitem[DeMiguel et~al.(2009)DeMiguel, Garlappi, and
  Uppal]{demiguel2009optimal}
V.~DeMiguel, L.~Garlappi, and R.~Uppal.
\newblock Optimal versus naive diversification: How inefficient is the 1/n
  portfolio strategy?
\newblock \emph{Review of Financial Studies}, 22\penalty0 (5):\penalty0
  1915--1953, 2009.

\bibitem[Feldman(2013)]{feldman2013techniques}
R.~Feldman.
\newblock Techniques and applications for sentiment analysis.
\newblock \emph{Communications of the ACM}, 56\penalty0 (4):\penalty0 82--89,
  2013.

\bibitem[Gottschalk and Gleser(1969)]{gottschalk1969}
L.~A. Gottschalk and G.~C. Gleser.
\newblock \emph{Measurement of Psychological States Through the Content
  Analysis of Verbal Behaviour}.
\newblock University of California Press, 1969.

\bibitem[Lipinski and Korczak(2004)]{Lipinski2004}
P.~Lipinski and J.~J. Korczak.
\newblock Performance measures in an evolutionary stock trading expert system.
\newblock \emph{Lecture Notes in Computer Science}, 3039:\penalty0 835--842,
  2004.

\bibitem[Markowitz(1952)]{markowitz1952portfolio}
H.~Markowitz.
\newblock Portfolio selection.
\newblock \emph{The Journal of Finance}, 7\penalty0 (1):\penalty0 77--91, 1952.

\bibitem[Oliveira et~al.(2013)Oliveira, Cortez, and
  Areal]{oliveira2013predictability}
N.~Oliveira, P.~Cortez, and N.~Areal.
\newblock On the predictability of stock market behavior using {S}tock{T}wits
  sentiment and posting volume.
\newblock \emph{Lecture Notes in Computer Science}, 8154:\penalty0 355--365,
  2013.

\bibitem[Oliveira et~al.(2014)Oliveira, Cortez, and
  Areal]{oliveira2014automatic}
N.~Oliveira, P.~Cortez, and N.~Areal.
\newblock Automatic creation of stock market lexicons for sentiment analysis
  using {S}tock{T}wits data.
\newblock In \emph{Proceedings of the 18th International Database Engineering
  \& Applications Symposium}, pages 115--123. ACM, 2014.

\bibitem[Peterson and Carl(2014)]{peterson2014}
B.~G. Peterson and P.~Carl.
\newblock \emph{PerformanceAnalytics: Econometric tools for performance and
  risk analysis}, 2014.
\newblock URL \url{http://CRAN.R-project.org/package=PerformanceAnalytics}.
\newblock R package version 1.4.3541.

\bibitem[{R Core Team}(2014)]{R2014}
{R Core Team}.
\newblock \emph{R: A Language and Environment for Statistical Computing}.
\newblock R Foundation for Statistical Computing, Vienna, Austria, 2014.
\newblock URL \url{http://www.R-project.org}.

\bibitem[Sharpe(1994)]{sharpe1994sharpe}
W.~F. Sharpe.
\newblock The sharpe ratio.
\newblock \emph{The Journal of Portfolio Management}, 21\penalty0 (1):\penalty0
  49--58, 1994.

\bibitem[Smailovi{\'c} et~al.(2014)Smailovi{\'c}, Gr{\v{c}}ar, Lavra{\v{c}},
  and {\v{Z}}nidar{\v{s}}i{\v{c}}]{smailovic2014stream}
J.~Smailovi{\'c}, M.~Gr{\v{c}}ar, N.~Lavra{\v{c}}, and
  M.~{\v{Z}}nidar{\v{s}}i{\v{c}}.
\newblock Stream-based active learning for sentiment analysis in the financial
  domain.
\newblock \emph{Information Sciences}, 285:\penalty0 181--203, 2014.

\bibitem[Surowiecki(2005)]{Surowiecki2005}
J.~Surowiecki.
\newblock \emph{The Wisdom of Crowds}.
\newblock Anchor Books, 2005.

\bibitem[Tumarkin and Whitelaw(2001)]{tumarkin2001news}
R.~Tumarkin and R.~F. Whitelaw.
\newblock News or noise? {I}nternet postings and stock prices.
\newblock \emph{Financial Analysts Journal}, 57\penalty0 (3):\penalty0 41--51,
  2001.

\end{thebibliography}
\end{document}